\documentclass[12pt]{article}
\usepackage[dvips]{graphicx}
\usepackage{float}
\usepackage{latexsym}
\usepackage{amsmath,mathrsfs,bm,amssymb,color}

\def\one{1}

\newtheorem{theorem}{Theorem}[section]
\newtheorem{proposition}[theorem]{Proposition}
\newtheorem{lemma}[theorem]{Lemma}
\newtheorem{corollary}[theorem]{Corollary}
\newtheorem{definition}[theorem]{Definition}

\newtheorem{remark}[theorem]{Remark}

\newtheorem{assumption}[theorem]{Assumption}

\newcommand{\eq}[1]{\begin{equation}\label{#1}}
\newcommand{\en}{\end{equation}}
\newcommand{\eqn}{\begin{eqnarray*}}
\newcommand{\enn}{\end{eqnarray*}}
\newcommand{\eqnn}{\begin{eqnarray}}
\newcommand{\ennn}{\end{eqnarray}}
\newcommand{\proof}[1][Proof]{{\sc #1.} }
\newcommand{\qed}{\hfill {\bf qed}\par\medskip}
\newcommand{\bi}{\begin{description}}
\newcommand{\ei}{\end{description} }
\newcommand{\bern}{\Psi}
\newcommand{\TTT}{\theta}
\newcommand{\OO}{{\ms W}}
\newcommand{\kkk}{{\rm P}}

\renewcommand{\c}{\TTT}
\newcommand{\tr}{\tau_R}

\newcommand{\CCC}{C_0^\infty}

\newcommand{\lkkk }{\left[}
\newcommand{\rkkk}{\right]}

\newcommand{\loc}[1]{L_{\rm loc}^{#1}(\RR^3)}

\newcommand{\VS}{{\bf \s}}

\newcommand{\bl}[1]{\begin{lemma}\label{#1}}
\newcommand{\el}{\end{lemma}}
\newcommand{\bc}[1]{\begin{corollary}\label{#1}}
\newcommand{\ec}{\end{corollary}}
\newcommand{\bt}[1]{\begin{theorem}\label{#1}}
\newcommand{\et}{\end{theorem}}
\newcommand{\bp}[1]{\begin{proposition}\label{#1}}
\newcommand{\ep}{\end{proposition}}
\newcommand{\br}[1]{\begin{remark}\label{#1}}
\newcommand{\er}{\end{remark}}
\newcommand{\bd}[1]{\begin{definition}\label{\rm #1}}
\newcommand{\ed}{\end{definition}}
\newcommand{\HH}{H}
\newcommand{\HHH}{H_{{\mathbb Z}_2}}
\newcommand{\HHHH}{H_\varepsilon }
\newcommand{\HHHHH}{H_{\rm NR}}

\newcommand{\hhh}{h_{{\mathbb Z}_2}}

\newcommand{\E}{{\mathbb E}}
\newcommand{\EE}{{\mathbb E}}

\newcommand{\pro}[1]{(#1_t)_{t\geq0}}

\newcommand{\CC}{{{\mathbb  C}}}
\newcommand{\U}{{{\rm F}}}
\newcommand{\RR}{{\mathbb  R}}
\newcommand{\BR}{{{\mathbb  R}^3 }}

\newcommand{\limn}{\lim_{n\rightarrow\infty}}

\newcommand{\SSS}{{\ms S}}
\newcommand{\SSSS}{{\ms S}_{\rm \!\! NR}}
\newcommand{\fff}{\ms F}
\newcommand{\mmmm}{\ms M}

\newcommand{\kak}[1]{(\ref{#1})}
\newcommand{\LR}{{L^2(\BR)}}
\newcommand{\LRR}{{L^2(\BR\times\zz)}}
\newcommand{\zz}{{\mathbb Z}_2}
\newcommand{\UI}{|U|}
\newcommand{\dv}{\, \dot +\,  V_+\, \dot - \, V_-}

\newcommand{\is}[1]{\inf{\rm Spec}({#1})}

\newcommand{\MMMM}{\EE_{M}^{x,\alpha,0}}
\newcommand{\MMMMM}{\EE_{M}^{0,0,0}}
\newcommand{\EEN}{\EE_{M}}

\newcommand{\AAA}{\sum_{\alpha=0,1}}

\newcommand{\ixx}{\sum_{\alpha=0,1}\int_{\RR^3} dx
\MMMM}
\newcommand{\lk}{\left(}
\newcommand{\rk}{\right)}
\newcommand{\lkk}{\left\{}
\newcommand{\rkk}{\right\}}
\newcommand{\ms}[1]{\mathscr{#1}}
\newcommand{\la}{\lambda }
\newcommand{\ov}[1]{\overline{#1}}
\newcommand{\gr}{\varphi_{\rm g}}

\newcommand{\mmm}[4]
{\left[ \!\!\!\begin{array}{cc}#1&#2\\
#3&#4\end{array}\!\!\!\right]}
\newcommand{\f}{^{-1}}
\newcommand{\hz}{h_0}

\newcommand{\half}{\frac{1}{2}}
\newcommand{\han}{{1/2}}

\newcommand{\ud}{U_{\rm d}}
\newcommand{\uo}{U_{\rm od}}

\newcommand{\mass}{m_\ast}

\newcommand{\vvv}[1]
{\lkkk \!\!\!\begin{array}{c}#1\end{array}\!\!\!\rkkk}

\newcommand{\MM}[4]
{\lkkk \!\!\!\begin{array}{cc}#1&#2\\ #3&#4\end{array}\!\!\!\rkkk}

\newcommand{\s}{\sigma}
\newcommand{\spinless}{H_{\rm spinless}}

\renewcommand{\d}{\displaystyle}
\newcommand{\non}{\nonumber}

\makeatletter
\@addtoreset{equation}{section}
\makeatother

\title
{\bf \large
Probabilistic Representation and Fall-Off of Bound States of
Relativistic Schr\"odinger Operators with Spin 1/2 }
\author{
 Fumio Hiroshima\thanks{
  Faculty of Mathematics, Kyushu University,
 Fukuoka, 819-0395,  Japan.
     hiroshima@math.kyushu-u.ac.jp.}, 
      Takashi Ichinose\thanks {
       Department of Mathematics,
Kanazawa University,
 Kanazawa, 920-11, Japan.
  ichinose@kenroku.kanazawa-u.ac.jp}  
   and J\'ozsef L\H{o}rinczi\thanks{
    School of Mathematics, Loughborough University,
     Loughborough LE11 3TU, United Kingdom.
       J.Lorinczi@lboro.ac.uk}}

\date{}

\begin{document}
\pagestyle{myheadings}
\markboth{Fall-off of bound states}
{Fall-off of bound states}
\maketitle
\setlength{\baselineskip}{14pt}

\begin{abstract}
\noindent
A Feynman-Kac type formula of relativistic Schr\"odinger operators with unbounded vector potential and spin
$1/2$ is given in terms of a three-component process consisting of Brownian motion, a Poisson process and a
subordinator. This formula is obtained for unbounded magnetic fields and magnetic fields with zeros. From this
formula an energy comparison inequality is derived. Spatial decay of bound states is established separately
for growing and decaying potentials by using martingale methods.
\end{abstract}
\bigskip

\newpage

\section{Introduction}
In the paper \cite{hil09} we constructed a Feynman-Kac formula for a generalized Schr\"odinger operator
with spin of the form
\eq{m1}
\bern(h(a,\s))+V.
\en
Here $V$ is a real-valued external potential, $\bern$ is an arbitrary Bernstein function with
$\bern(0)=0$, and $h$ is a Schr\"odinger-type operator of the form
\eq{s1}
h(a,\s) =\half(\VS  \cdot (p-a))^2,
\en
including a vector potential $a=(a_1,a_2,a_3)$ describing a magnetic field, and the Pauli matrices
$\VS=(\VS_1,\VS_2,\VS_3)$ describing  spin $\han$. As we have shown, the Feynman-Kac representation
of \kak{m1} involves three independent stochastic processes, Brownian motion, a Poisson process and
a subordinator. Moreover, spin $\han$ was also extended to higher spins in \cite{hil09}, see also
\cite{ars91}.

In this paper we consider a functional integral representation of the strongly continuous one-parameter
semigroup generated by the relativistic Schr\"odinger operator with spin $1/2$ in three-dimensional
space,
\eq{m11}
\sqrt{(\VS\cdot(p-a))^2+m^2}-m+V.
\en
Here $m$ is the mass of the relativistic particle, which we regard as a parameter; see \cite{car78} 
for standard Schr\"odinger operators, where $a=0$. This Hamilton operator is a special case of \kak{m1} 
obtained by choosing
\eq{g3}
\bern(u)=\sqrt{2u+m^2}-m,\quad m\geq0.
\en
In this case we have the $\frac{1}{2}$-stable subordinator about which more details are known than
about subordinators related to a general $\Psi$. Using this extra information, our main goal in this
paper is to prove a Feynman-Kac-type formula for \kak{m11} under weaker conditions than needed for
general $\Psi$, and use it to derive the fall-off properties of bound states. In particular, in
contrast to \cite{hil09} we can cover {unbounded} magnetic fields in Theorem \ref{main} and magnetic
fields with zeros in Theorem \ref{main2}.

This paper is organized as follows. Section 2 is devoted to introducing the relativistic Schr\"odinger
operator with spin $\han$  as a self-adjoint operator on $\CC^2\otimes\LR$ and a unitary equivalent
representation on $\LRR$. In Section 3.1 we reassess results in \cite{hil09} and give a Feynman-Kac
formula with bounded magnetic fields. In Section 3.2 we prove a Feynman-Kac formula for unbounded
magnetic fields, and in Section 3.3 for magnetic fields having zeros. In Section 4 we derive the
decay properties of bound states separately for growing and decaying potentials by using martingale
methods.

\section{Relativistic Schr\"odinger operator with spin $\han$}
\subsection{Definitions}
We begin by defining the self-adjoint operator $h(a,\s)$ and $\sqrt{2h(a,\s)+m^2}-m+V$ rigorously.

The spinless Schr\"odinger operator $\hz$ with vector potential $a$ and zero external potential
is defined as a self-adjoint operator on  $\LR$. Let ${\rm D}_\mu=p_\mu-a_\mu$, where $p_\mu=
-i\partial_{x_\mu}$ is the generalized differential operator. Define the quadratic form $q$ by
\eq{y100}
H^1(\BR)\times H^1(\BR)\ni (f,g)\mapsto q (f,g)=\half \sum_{\mu=1}^3({\rm D}_\mu f,  {\rm D}_\mu g),
\en
where $H^1(\BR)=\{f\in \LR\,|\, {\rm D}_\mu f\in \LR,\, \mu=1,2,3\}$. If $a\in (\loc 2)^3$, then the
quadratic form $q$ is non-negative and closed, and hence there exists a unique self-adjoint operator
$\hz$ satisfying $(\hz f,g)=q (f,g)$, for $f\in D(\hz )$ and $g\in H^1$, where $D(\hz )=
\left\{f\in {\rm Q}(q )\,|\,q (f,\cdot)\in \LR'\right\}$. Let $C_0^\infty(\BR)=\CCC$ be the set of
infinitely many times differentiable functions with compact support on $\BR$. It can be seen that
$C_0^\infty$ is a form core for $\hz$ under the assumption $a\in (\loc 2)^3 $, see \cite{ls81}.

Next we introduce a magnetic field $b=(b_1,b_2,b_3)$. Physically it is given by $b=\nabla\times a$,
however, in this paper we regard the magnetic field $b$ independent of the vector potential $a$. We
will use the following conditions on the vector potential $a$.
\begin{assumption}[\textbf{Vector potential}]
\label{fumio4}
{\rm
The vector potential $a=(a_1,a_2,a_3)$ is a vector-valued function whose components $a_\mu$, $\mu
= 1,2,3$, are real-valued functions such that $a\in (\loc 2)^3 $ and $\nabla\cdot a\in \loc 1$,
where $\nabla\cdot a$ is understood in distributional sense.
}
\end{assumption}
\begin{assumption}[\textbf{Magnetic field}]
\label{70}
Suppose that $D(-\Delta)\subset D(b_\mu)$ and for $f\in D(-\Delta)$ the conditions $\|b_\mu f\|\leq
\kappa_\mu \|-\Delta f\|+\kappa_\mu'\|f\|$, $\mu = 1,2,3$, and $\kappa_1+\kappa_2+\kappa_3<1$ are
satisfied.
\end{assumption}

Finally we introduce the spin variables. Let $\VS =(\s_1,\s_2,\s_3)$ be the $2\times 2$ Pauli matrices
given by
$$
\s_1=  \mmm 0 1 1 0, \quad \s_2= \mmm 0 {-i} i 0 , \quad \s_3= \mmm 1 0 0 {-1}.
$$
They satisfy the relations $\s_\mu\s_\nu+\s_\nu\s_\mu  =2\delta_{\mu\nu}1$ and $\s_\mu\s_\nu=
i\sum_{\la =1}^3 \varepsilon^{\la \mu\nu}\s_\la $, where $\varepsilon^{\la \mu\nu}$ is the
anti-symmetric Levi-Civit\`a tensor with $\varepsilon^{123}=1$. Then it can be seen directly that
$$
\s\otimes  b=\sum_{\mu=1}^3\sigma_\mu\otimes b_\mu=\MM{b_3} {b_1-ib_2}{b_1+b_2} {-b_3}.
$$
Under Assumption \ref{70} $\s\otimes b$  is relatively bounded with respect to $\one \otimes
2h_0$, as an operator in $\CC^2\otimes\LR$,  with a relative bound strictly smaller than 1,
\eq{kubo}
\|(\s\otimes  b) f\|\leq (\kappa_1+\kappa_2+\kappa_3)\|\one\otimes 2h_0 f\|+C\|f\|, \quad
f\in \CC^2\otimes D(h).
\en
This follows through the diamagnetic inequality $|(f, e^{-t\hz}g)| \leq
(|f|, e^{-t (-\half\Delta)}|g|)$ under Assumption \ref{fumio4}. Thus the self-adjoint operator
\eq{ss1}
h=\one\otimes \hz-\half \s\otimes b
\en
in $\CC^2\otimes\LR$ is bounded from below under Assumptions \ref{fumio4} and \ref{70}.
We choose $m$ so as to guarantee that
$$
2h+m^2=\one\otimes 2h_0-\s\otimes b+m^2\geq 0.
$$
Note that under a suitable condition $h$ is positive, and in this case we can take $m=0$. From
now on we omit the tensor product $\otimes$ for notational convenience.

We now define the self-adjoint operator $H$.
\begin{definition}
{\rm
Under Assumptions  \ref{fumio4} and \ref{70},  $H$ is defined by the self-adjoint operator
\eq{def}
H=\sqrt{2h+m^2} -m
\en
in $\CC^2\otimes\LR$.
Here the square root is taken through the spectral resolution of $2h+m^2$.
}
\end{definition}
An example is the operator $\sqrt{(\sigma\cdot(p-a))^2+m^2}-m$ such that $a\in (L_{\rm loc}^4(\BR))^3$,
$\nabla\cdot a\in L_{\rm loc}^2(\BR)$ and $\nabla\times a\in (L_{\rm loc}^2(\BR))^3$. In this case it
is seen that
$$
(\sigma\cdot(p-a))^2= (p-a)^2+\s\cdot  (\nabla\times a)
$$
on $\one\otimes C_0^\infty(\BR)$.

\subsection{Spin variable}
In order to construct a functional integral representation of $(f, e^{-t(H+V)}g)$ we make a unitary
transform of $H$ on $\CC^2\otimes\LR$ to an operator on the space
$L^2({\mathbb  R}^3 \times {\mathbb  Z}_2)$. This is a space of $L^2$-functions of $x \in \RR^3$ and
an additional two-valued spin variable $\c \in {\mathbb  Z}_2$,
where
\eq{ko25}
\zz= \{-1,1\}.
\en
We define the spin interaction $U$ on $L^2(\RR^3\times \zz )$ by
\eq{u}
U:f(x,\c)\mapsto \ud(x,\c)f(x,\c)+\uo(x,-\c)f(x,-\c)
\en
where $(x,\c)\in\BR\times\zz$,
\eq{hh}
\ud(x,\c)=-\half \c b_3(x)
\en
is the diagonal component, and
\eq{hh2}
\uo (x,-\c)=-\half (b_1(x)-i\c b_2(x))
\en
is the off-diagonal component. Let
\eq{yasumi}
\hhh=h_0+U.
\en
Under Assumption \ref{70} $U$ is symmetric,
relatively bounded with respect to $\hz$ with
a relative
bound strictly smaller than 1 so that $\hhh$ and $h$ are unitary equivalent,
\eq{eren}
\hhh\cong h
\en
as seen below. Define the unitary operator $\U:L^2(\RR^3\times \zz )\rightarrow \CC^2\otimes\LR$
by
\eq{uni2}
\U: f\mapsto \vvv {f(\cdot,+1)\\ f(\cdot,-1)}.
\en
Also, define $\tau_\mu=F^{-1} \s_\mu F$. We see that
$\tau_1: f(x,\theta)\mapsto f(x,-\theta)$,
$\tau_2: f(x,\theta)\mapsto -i\theta f(x,-\theta)$ and
$\tau_3: f(x,\theta)\mapsto \theta f(x,\theta)$.
\begin{definition}
{\rm
Let Assumptions \ref{fumio4}  and  \ref{70} hold. Then $\HHH$ is defined by
\eq{ko34}
\HHH =\sqrt{ 2\hhh +m^2}-m.
\en
}
\end{definition}

\noindent
In what follows instead of $\HH$ we study $\HHH$, and write $\HH$ (resp. $h$) instead of $\HHH$
(resp. $\hhh$).

\subsection{Three independent stochastic processes}
In order to construct a path integral representation we will need three independent stochastic
processes $\pro B$, $\pro N$ and $\pro T$ which we introduce next.
We denote the expectation with respect to path measure $W$ starting at $x$ by $\EE_W^x$.

Let $\pro B$ be three-dimensional Brownian motion on a probability space $(\Omega_P,\ms F_P,P^x)$
with initial point $P^x(B_0=x)=1$.

Secondly, let $\pro N$ be a Poisson process on a probability  space $(\Omega_N,\ms F_N, \mu)$ with
unit intensity, i.e.,
$$
\mu(N_t=n)=\frac{t^n}{n!}e^{-t},\quad n\in{\mathbb N}\cup\{0\}.
$$
Let $\mu^\alpha$ be the image measure of the process $(N_t+\alpha)_{t\geq 0}$ for $\alpha\in\zz$ and 
thus $\EE_\mu^\alpha[f(N_\cdot)]=\EE_\mu^0[f(N_\cdot+\alpha)]$. We define integrals with respect to this 
process in terms of the sum of evaluations at jumping times, i.e., for  $g$ we define
\eq{poisson}
\int_{(a,b]}g(s,N_{s})dN_s=\sum_{r\in (a,b] \atop N_{r+}\not= N_{r-}} g(r,N_{r})
\en
and we also write $\d\int_a^{b+}\cdots dN_s$ for $\d\int_{(a,b]}\cdots dN_s$. Associated with the Poisson 
process we also define a $\zz$-valued stochastic process $\pro \theta$ on $(\Omega_N,\ms F_N,\mu^0)$ by
\eq{hh6}
\c_t=(-1)^{N_t}.
\en

Finally, let $\pro T$ denote the subordinator starting from $0$ at $t=0$ on a given probability space 
$(\Omega_\nu,\ms F_\nu,\nu)$ defined by its Laplace transform
\eq{g5}
\EE_\nu^0[e^{-T_t u}]=\exp\lk -t\lk \sqrt{2u+m^2}-m\rk
\rk.
\en
Note that $\pro T$ is a one-dimensional L\'evy process with right continuous paths with left limits, almost
surely non-decreasing. It can be more explicitly described as the first hitting time process
$$
T_t=\inf\{s>0 \,|\, B_s^1+m s = t\},
$$
where $(B^1_t)_{t\geq 0}$ is a one-dimensional Brownian motion independent of the three-dimensional Brownian
motion $B_t$ above. We also define the  measure $\nu^s$, $s\in\RR$,  by the image measure on $(T_t+s)_{t\geq 0}$,
and use the shorthand
\eq{expectation}
\EE_P^x \EE_\mu^\alpha \EE_\nu^s =\EE^{x,\alpha,s}_{M}.
\en
The role of these three stochastic processes is as follows. Clearly, the Schr\"odinger operator $-\half\Delta+V$
generates an It\^o process which can be described using the Brownian motion $\pro B$ under $V$. The Poisson process
$\pro N$ results from the Schr\"odinger operator with spin. Finally, the subordinator $\pro T$ appears due to the
relativistic Schr\"odinger operator which generates a L\'evy process. A particular combination of these three
independent stochastic processes then yields the path integral representation of $e^{-t(\HH+V)}$ which we will
discuss below.

\subsection{Generator of Markov process}
Consider the  $\RR^3\times\zz$-valued joint Brownian and jump process
$$
\Omega_P\times \Omega_N \ni (\omega,\omega_1)\mapsto
X_t (\omega,\omega_1)=(B_t(\omega),\c_t(\omega_1)) \in\BR\times\zz
$$
with initial value $X_0$. The generator of this  Markov process is \cite{hil09}
\eq{gg}
G_0= -\half \Delta+\s_{\rm F}+\one,
\en
where $\one$ is the $2\times 2$ identity matrix and $\s_{\rm F}$ is the fermionic harmonic oscillator defined in 
terms of the Pauli matrices by
$$
\s_{\rm F}=\half (\s_3+i\s_2)(\s_3-i\s_2)- \one =-\s_1.
$$
Note that $\is {G_0}=0$.

In the relativistic case, the subordinator explained above appears in addition to this. We define the
subordinate process $\pro q$ in terms of the $\BR\times\zz$-valued stochastic process
$$
\Omega_P\times\Omega_N\times\Omega_\nu \ni (\omega,\omega_1,\omega_2)\mapsto
q_t(\omega,\omega_1,\omega_2)=(B_{T_t(\omega_2)}(\omega), \c_{T_t(\omega_2)}(\omega_1))\in \BR\times\zz.
$$
In a similar manner to $\pro X$, we can identify the generator of $\pro q$.
\begin{proposition}
The generator of the Markov process $\pro q$ is
\eq{g6}
G=\sqrt{-\Delta+2 \s_{\rm F}+2+m^2}-m
\en
and its characteristic function is given by
\eq{c10}
\EEN^{0,0,0}[e^{i Zq_t}]=\EEN^{0,0,0}[e^{i \xi B_{T_t}}e^{i z\theta_{T_t}}]=
e^{-t(\sqrt{|\xi|^2+m^2}-m)} \cos z + ie^{-t(\sqrt{|\xi|^2+4+m^2}-m)}\sin z
\en
for $Z=(\xi,z)\in \BR\times \RR$.
\end{proposition}
\proof
This is obtained through the equalities
\begin{eqnarray*}
\AAA \int_\BR dx \MMMM \left[\ov{f(q_0)}g(q_t) \right]
&=&
\EE_\nu^0\left[\AAA \int_\BR dx \EE^{x,\alpha}_{P\times\mu}
 \left[\ov{f(q_0)}g(q_t) \right]\right]\\
&=&
\EE_\nu^0[(f, e^{-T_t(-\half\Delta+\sigma_F+\one)}g)] =
(f, e^{-tG}g).
\end{eqnarray*}
Hence it follows that \kak{g6} is the generator of $\pro q$, while \kak{c10} is straightforward.

\qed

\section{Feynman-Kac-type representations}
\subsection{Bounded magnetic field}
In this subsection we briefly discuss some results established in \cite{hil09} obtained for a
general version of the relativistic Schr\"odinger operator with spin and bounded magnetic field.
Write
\eq{w}
W(x)=\half \sqrt{b_1(x)^2+b_2(x)^2},
\en
and notice that $|\uo(x,\theta)|=W(x)$.

\begin{proposition}[\textbf{Feynman-Kac formula: bounded magnetic field}]
\label{main0}
Let Assumption \ref{fumio4} hold and assume that $b_\mu\in L^\infty$ for $\mu = 1,2,3$. Let $V$ be
relatively bounded with respect to $\sqrt{-\Delta +m^2}$ with a relative bound strictly smaller
than 1. Assume, furthermore, that
\eq{hh3210}
\EE^{x,0}_{P\times\nu}
\lkkk\int_0^{T_t} {|\log W(B_s)|}ds\rkkk <\infty,\quad {\rm a.e.} \; x\in\RR^3.
\en
Then
$\HH+V$ is self-adjoint on $D(\HH)$ and
\eq{ko16}
(f, e^{-t(\HH  + V)}g)=\AAA\int_{\RR^3} dx \MMMM\lkkk e^{T_t}
\ov{f(q_0)} g(q_t)e^{\SSS} \rkkk,
\en
where  the exponent
$\SSS=\SSS_V +\SSS_{\rm A} +\SSS_{\rm S}$ is given by
\begin{align}
&
\label{london1}
\SSS_V  =-\int_0^t V(B_{T_s}) ds, \\
&
\label{london2}
\SSS_{\rm A}  =-i\int_0^{T_t } a(B_s)\circ dB_s,\\
&
\label{london3}
\SSS_{\rm S} = -\int_0^{T_t }
\ud
(B_s,\c_s) ds + \int_0^{T_t+} \log\lk -\uo(B_s,- \c_{s-})\rk dN_s.
\end{align}
\end{proposition}
\proof
Since $\|Vf\|\leq \kappa \|\sqrt{-\Delta+m^2}f\|+\kappa'\|f\|$ with constants $\kappa<1$ and
$\kappa'$, and $b_\mu$ is bounded, we have $\|Vf\| \leq \kappa \|\HH f\|+ C \|f\|$ with a
constant $C$. Hence self-adjointness follows by the Kato-Rellich theorem. \kak{ko16} follows
from \cite[Theorem 5.9]{hil09}.
\qed
We note  that $\SSS_{\rm A}$ and $\SSS_{\rm S}$ in Proposition \ref{main0}  stand for the 
integrals $-i\int_0^r a(B_s)\circ dB_s$ and $-\int_0^r \ud (B_s,\c_s) ds + 
\int_0^{r+} \log\lk - \uo(B_s,- \c_{s-})\rk dN_s$ evaluated at $r=T_t$, respectively.

A Feynman-Kac formula without spin is an immediate corollary. This was first established in
\cite{CMS90} without a vector potential; we give a version including a vector potential. Let
\eq{spinless}
\spinless=\sqrt{2h_0+m^2}-m.
\en
\bc{ko52}
Let Assumption \ref{fumio4} hold, and assume that $V=V_+-V_-$ satisfies that $V_+\in
L_{\rm loc}^1(\BR)$ and $V_-$ is relatively form bounded with respect to $\sqrt{-\Delta+m^2}$
with a relative bound strictly less than 1. Then
\eq{ko42}
(f, e^{-t(\spinless \dv)}g) =\int_{\BR} dx \EE_{P\times  \nu}^{x,0} \lkkk \ov{f(B_0)}
g(B_{T_t}) e^{\SSS_V +\SSS_{\rm A} } \rkkk.
\en
In particular, when $a=0$,
\eq{ko421}
(f, e^{-t(\sqrt{-\Delta+m^2}-m \dv)}g)=\int_{\BR} dx \EE_{P\times  \nu}^{x,0} \lkkk \ov{f(X_0)}
g(X_t) e^{-\int_0^tV(X_s) ds}\rkkk.
\en
\ec
By Corollary \ref{ko52} we have the following energy comparison inequality. Let
\eq{hzero}
\HH_0=\sqrt{-\Delta+m^2}-m.
\en
\bc{dia3}
Under the assumptions of Corollary
\ref{ko52} we have
\begin{enumerate}
\item[(1)]
$|(f, e^{-t(\spinless \dv )}g)| \leq (|f|, e^{-t(\HH_0 \dv )}|g|)$
\item[(2)]
$\is {\HH_0 \dv }\leq \is {\spinless \dv }$.
\end{enumerate}
\end{corollary}

\subsection{Unbounded magnetic field}
We extend the Feynman-Kac formula above (Proposition \ref{main0}) to the case of magnetic fields $b$
that are possibly unbounded and satisfy Assumption \ref{70}. This extension is not straightforward,
and we need several lemmas.

Define the truncated magnetic field $b^{(N)}$ by
$$
b_\mu^{(N)}(x) = \lkk
\begin{array}{ll}
b_\mu(x) & \mbox{if} \;\; |b_\mu(x)|\leq N \\
N & \mbox{if} \;\; b_\mu(x)>N \\
-N & \mbox{if} \;\; b_\mu(x)<-N.
\end{array}
\right.
$$
Then the Feynman-Kac formula for the Hamiltonian with the truncated magnetic field is readily given
by Proposition \ref{main0} in which $b$ is replaced by $b^{(N)}$. Let $\HH_N$ be defined by $\HH$
with $b$ replaced by $b^{(N)}$.

\bl{n2}
Under Assumptions \ref{fumio4} and \ref{70} the semigroup  $e^{-t\HH_N}$ is strongly convergent
to $e^{-t\HH}$ as $N\to\infty$.
\el
\proof
Let $h_N$ be $h$ with $b$ replaced by $b^{(N)}$. We see that $h_N\to h$ as $N\to\infty$ on the common
domain $D(h_n)=D(h)$. Then $e^{-th_N}\to e^{-th}$ strongly as $N\to\infty$. Thus it is immediate to
see that
\eq{ha1}
(f, e^{-t\HH_N}g)=\EE_\nu^0[(f, e^{-T_t h_N }g)]\to \EE_\nu^0[(f, e^{-T_t h}g)]=(f, e^{-t\HH}g),
\en
which implies strong convergence.
\qed
\bl{n3}
Let $f,g\in\LRR$, and  set
$$
\rho=f(q_0)g(q_t) e^{\int_0^{T_t}\half |b_3(B_s)|ds}
e^{\int_0^{T_t+}\log  W(B_s) dN_s}e^{T_t}.
$$
Then
under Assumption \ref{70}
it follows that
 $\d \AAA\int_{\RR^3} dx\MMMM[|\rho|]<\infty$.
\el
\proof
Define the spin operator $\UI$ and $\UI_N$ by
\begin{align}
&\UI: f(x,\theta)\mapsto -\half |b_3(x)|f(x,\theta)-W(x)f(x,-\theta),\\
&\UI_N: f(x,\theta)\mapsto -\half |b_3^{(N)}(x)|f(x,\theta)-W^{(N)}(x)f(x,-\theta),
\end{align}
where $W^{(N)}$ is $W$ with $b$ replaced by $b^{(N)}$,
 and define
\eq{n1}
\widehat \HH=\sqrt{-\Delta+2\UI+m^2}-m.
\en
Also, we define $\widehat\HH_N$ by $\widehat \HH$ with $\UI$ replaced by $\UI_N$. Let
$f,g\in\LR$ be non-negative. For $\widehat \HH_N$ we have the Feynman-Kac formula
\eq{ko16-1}
(f, e^{-t\widehat \HH_N}g)=\AAA\int_{\RR^3} dx \MMMM\lkkk e^{T_t}
\ov{f(q_0)} g(q_t)e^{\widehat\SSS_{\rm S}^{N}} \rkkk,
\en
where
\eq{mo1}
\widehat\SSS_{\rm S}^{N}=\int_0^{T_t}\half |b_3^{(N)}(B_s)| ds+
\int_0^{T_t+}\log  W^{(N)}(B_s) dN_s.
\en
By the monotone convergence theorem for forms we see that $e^{-t(-\Delta+2\UI_N)}\to e^{-t(-\Delta+2\UI)}$
strongly as $N\to\infty$, and thus $e^{-t\widehat\HH_N}\to e^{-t\widehat\HH}$ strongly as $N\to\infty$ is
shown in the same way as \kak{ha1}. Then the monotone convergence theorem for integrals implies that $\rho$
is integrable and the Feynman-Kac formula \kak{ko16-1} with $b^{(N)}$ replaced by $b$ also holds.
\qed

Now we can state the first main theorem.
\begin{theorem}
[\textbf{Feynman-Kac formula: unbounded magnetic field}]
\label{main}
Let Assumptions \ref{fumio4} and \ref{70} as well as condition \kak{hh3210} hold, and suppose that $V$
is relatively bounded with respect to $\sqrt{-\Delta +m^2}$ with a relative bound strictly less than 1.
Then $H+V$ is self-adjoint on $D(H)$ and
\eq{MAIN}
(f, e^{-t(\HH  + V)}g)=\AAA\int_{\RR^3} dx \MMMM\lkkk e^{T_t}
\ov{f(q_0)} g(q_t)e^{\SSS} \rkkk.
\en
\end{theorem}
\proof
We divide the proof in five steps.

\medskip
\noindent
{\bf \emph{Step 1}:} Suppose that $V=0$. Then the theorem holds.

\vspace{0.1cm}
\noindent
\emph{Proof:}
Recall that $\HH_N$ is defined by $\HH$ with $b$ replaced by $b^{(N)}$. Then the Feynman-Kac formula
holds with $\SSS_{\rm S}$ replaced by $\SSS_{\rm S}^N$, where $\SSS_{\rm S}^N$ is defined by
$\SSS_{\rm S}$ with $b$ replaced by $b^{(N)}$:
\eq{ma1}
(f, e^{-t\HH_N}g)=\AAA\int_{\RR^3} dx\MMMM\left[e^{T_t}\ov{f(q_0)}g(q_t)
e^{\SSS_{\rm S}^N+\SSS_{\rm A}}\right].
\en
The left hand side above converges to $(f, e^{-t\HH}g)$ as $N\to \infty$ by Lemma \ref{n2}. On the
other hand, we have
$$
e^{T_t}|f(q_0)g(q_t)||e^{\SSS_{\rm S}^N+\SSS_{\rm A}}| \leq
e^{T_t}|f(q_0)g(q_t)| e^{\int_0^{T_t}\half |b_3(B_s)|ds}
e^{\int_0^{T_t+}\log  W(B_s) dN_s}
$$
so that the right hand side of \kak{ma1} is integrable by Lemma \ref{n3}, and therefore the Lebesgue
dominated convergence theorem yields
$$
\lim_{N\to\infty}\AAA \int_{\RR^3} dx\MMMM
\left[e^{T_t}f(q_0)g(q_t)e^{\SSS_{\rm S}^N+\SSS_{\rm A}}\right]=
\AAA\int_{\RR^3} dx\MMMM\left[e^{T_t}f(q_0)g(q_t)e^{\SSS_{\rm S}+\SSS_{\rm A}}
\right].
$$
Hence the theorem follows for $V=0$.

\medskip
\noindent
{\bf \emph{Step 2:}} $V$ is relatively bounded with respect to $H$ with
a relative bound strictly
smaller than 1. In particular, $H+V$ is self-adjoint on $D(H)$.

\vspace{0.1cm}
\noindent
\emph{Proof:} Let $b_0=(\sqrt{b_1^2+b_2^2},0,b_3)$ and $\HH_{b_0}$ be defined by $\HH$ with
$a=0$ and $b$ replaced by $b_0$, i.e., $\HH_{b_0}=\sqrt{-\Delta+\s\cdot b_0+m^2}-m$.
Set $\s\cdot b_0=U_{b_0}$. Then we have
$$
\|\sqrt{-\Delta +m^2}f\|^2=\|(\HH_{b_0}+m)f\|^2 +(f, -U_{b_0}f).
$$
Since $|(f, U_{b_0} f)|\leq \kappa' \|f\|^2$ with a constant $\kappa'$, and  $\|Vf\|\leq
\kappa \|\sqrt{-\Delta+m^2}f\|+\kappa''\|f\|$ with constants $\kappa<1$ and $\kappa''$, we have
$\|Vf\| \leq A \|\HH_{b_0}f\| + C\|f\|$ with some $C$ and $A<1$. From the Feynman-Kac formula
established in Step 1 the diamagnetic inequality,
\eq{dia}
|(f, e^{-t\HH}g)|\leq (|f|, e^{-tH_{b_0}}|g|)
\en
follows. From \kak{dia}  we have
$\|\HH_{b_0}f\|\leq\|\HH f\|+c\|f\|$, and thus
$$
\|Vf\| \leq A \|\HH f\|+ C' \|f\|
$$
with a constant $C'$.  Hence self-adjointness follows by the Kato-Rellich theorem.

\medskip
\noindent
{\bf\emph{Step 3:}} Suppose $V\in L^\infty(\BR)\cap C(\BR)$. Then the statement holds.

\vspace{0.1cm}
\noindent
\emph{Proof:} By the Trotter product formula and the Markov property of $\pro q$ we have
that
\begin{eqnarray*}
(f, e^{-t(\HH+V)}g)
&=&
\limn (f, (e^{-(t/n)\HH}e^{-(t/n) V})^n g)\\
&=&\AAA\int_{\RR^3} dx\MMMM
\left[e^{T_t} f(q_0)g(q_t) e^{-\sum_{j=1}^n (t/n)V(B_{T_{tj/n}})} e^{\SSS_{\rm S}+\SSS_{\rm A}}\right].
\end{eqnarray*}
Note that $s\mapsto V(B_{T_s})$ is continuous in $s\in [0,t]$ except for at most finitely many points.
Thus
$$
-\sum_{j=1}^n (t/n)V(B_{T_{tj/n}(\omega_2)}(\omega))\stackrel {n\to\infty}\to
-\int_0^t V(B_{T_s(\omega_2)}(\omega))ds
$$
for almost every $(\omega,\omega_2)\in\Omega_P\times\Omega_\nu$ as a Riemann integral. Then the
theorem follows for $V\in L^\infty(\BR)\cap C(\BR)$.

\medskip
\noindent
{\bf \emph{Step 4:}} Suppose $V\in L^\infty(\BR)$. Then the statement holds.

\vspace{0.1cm}
\noindent
\emph{Proof:}
Let  $V_n=\phi(\cdot/n)(V\ast j_n)$, where $j_n(x)=n^3\phi(xn)$ with $\phi\in C_0^\infty$ such that
$0\leq \phi\leq 1$, $\int_\BR \phi(x) dx=1$ and $\phi(0)=1$. Then $V_n(x)\to V(x)$for $x\not\in \ms N$,
where $\ms N$ is a set of Lebesgue measure zero. Notice that
$$
\EE_{P\times \nu}^{x,0}[1_{\ms N}(B_{T_s})] = \int_\BR 1_{\ms N}(x+y)\kkk_s(y) dy =0
$$
for $x\in \ms N$, where
$$
\kkk_s(x)=2\lk\frac{m}{2\pi}\rk^2 \frac{s}{s^2+|x|^2}K_2\lk m\sqrt{|x|^2+s^2}\rk
$$
is the distribution of the random variable $B_{T_s}$ and
$$\d K_2(x)=\half \int_0^\infty
\xi e^{-\half(\xi+\xi\f)x}d\xi$$ is the modified Bessel function of the third kind. Hence
$$
0=\int _0^t \E _{P\times\nu}^{x,0}[1_{\ms N}(B_{T_s})]ds=\E _{P\times\nu}^{x,0}\left[\int _0^t 1_{\ms N}(B_{T_s})ds\right].
$$
Then the Lebesgue measure of $\{s\in [0,\infty)\,|\,B_{T_s(\omega_2)}(\omega)\in \ms N\}$ is zero
for almost every path $(\omega,\omega_2)\in\Omega_P\times\Omega_\nu$. Therefore $\int_0^t V_n(B_{T_s})ds
\rightarrow \int_0^t V(B_{T_s})ds$ as $n\rightarrow \infty$ for almost every path $(\omega,\omega_2)\in
\Omega_P\times\Omega_\nu$. Moreover,
\begin{eqnarray*}
\lefteqn{
\ixx \left[e^{T_t}\ov{f(q_0)}g(q_t) e^{\SSS_{\rm A}+\SSS_{\rm S}}
e^{-\int_0^t V_n(B_{s})ds} \right]}\\
&&
\stackrel {n\to\infty} \rightarrow \ixx \left[e^{T_t}\ov{f(q_0)}g(q_t) e^{\SSS_{\rm A}+\SSS_{\rm S}}
e^{-\int_0^t V(B_{s})ds} \right].
\end{eqnarray*}
On the other hand, $e^{-t(H+V_n)}\rightarrow e^{-t(H+V)}$ strongly as $n\rightarrow \infty$, since
$H+V_n$ converges to $H+V$ on the common domain $D(H)$. Then the theorem follows for $V\in L^\infty(\BR)$.

\medskip
\noindent
{\bf \emph{Step 5:}} We complete the proof of Theorem \ref{main}. Let $V=V_+-V_-$ and $V_{mn}=V_{+m}-V_{-n}$,
with $V_+$, $V_-$ denoting the positive and negative parts of $V$, respectively, and $V_{+m}(x)=V_+(x)$ if
$V_+(x)\leq m$, and $V_+(x)=m$ if $V_+(x)\geq m$, similarly $V_{-n}(x)=V_-(x)$ if $V_-(x)\leq n$ and
$V_-(x)=n$ if $V_-(x)\geq n$. Then by the monotone convergence theorem for forms, we have
$e^{-t(H+V_{mn})}$ strongly converges to $e^{-t(H+V_{m\infty})}$ as $n\to \infty$, and furthermore
$e^{-t(H+V_{m\infty})}$ strongly converges to $e^{-t(H+V)}$ as $m\to\infty$. Hence
$$
\lim_{m\to\infty} \limn e^{-t(H+V_{mn})}=e^{-t(H+V)}.
$$
On the other hand, by the monotone convergence theorem for integrals the right hand side converges.
This completes the proof of the theorem.
\qed

\subsection{Magnetic field with zeros}
Next we consider the case when the off-diagonal component  $\uo(x,-\c)$ vanishes for some $x\in \BR$. In
this case it is not clear whether $\int_0^{t+}|\log W(B_s)|dN_s<\infty$ holds almost surely. An example
when this is not the case is obtained by choosing $b\in (\CCC )^3$.

Let
$$
\delta_\varepsilon(z)= \lkk
\begin{array}{ll} 1,&|z|<\varepsilon,\\
0,&|z|\geq \varepsilon,
\end{array}
\right.
$$
for $z\in\CC$ and write
\eq{sa1}
\chi_\varepsilon (z)=z+\varepsilon \delta_\varepsilon(z),\quad z\in\CC.
\en
We see that
$$
\left|\chi_\varepsilon \lk \uo(x,-\c) \rk\right| >\varepsilon,\quad
(x,\c)\in \RR^3\times\zz.
$$
Define $h_\varepsilon$ by $h $ with the off-diagonal part
replaced by $\chi_\varepsilon\lk \uo(x,-\c)\rk$,
i.e.,
$$
h_\varepsilon f(x,\c) = \lk \hz+\ud(x,\c) \rk f(x,\c )+\chi_\varepsilon \lk
\uo(x,-\c)\rk f(x,-\c),\quad
(x,\c)\in\BR\times\zz.
$$
Also, define $\HH_\varepsilon $ by $H$ with $\uo$ replaced by $\chi_\varepsilon \lk \uo(x,-\c)\rk$.

We note that for every $(x,\omega,\omega_1,\omega_2)\in \BR\times \Omega_P\times
\Omega_N\times\Omega_\nu$ there exists a number $n=n(\omega_1,\omega_2)$ and random jump times
$r_1(\omega_1),\ldots,r_{n}(\omega_1)$ of $s\mapsto N_s$ for $0\leq s\leq T_t(\omega_2)$ such that
\begin{eqnarray*}
\int_0^{T_t(\omega_2)+} \log W(x+B_s(\omega))dN_s =\sum_{j=1}^{n(\omega_1,\omega_2)}
\log W(x+B_{r_j(\omega_1)} (\omega)).
\end{eqnarray*}
Consider
\eq{oo}
\OO=\lkk (x,\omega,\omega_1,\omega_2) \in \BR\times\Omega_P\times \Omega_N\times\Omega_\nu \left|
\int_0^{T_t+}\log W(x+B_s) dN_s>-\infty \right.\rkk.
\en
Notice that by the definition $(x,\omega,\omega_1,\omega_2)\in\OO^c$ if and only if there exists $r$
such that
\begin{enumerate}
\item[(1)] $0< r\leq t\leq T_t(\omega_2)$,
\item[(2)] $s\mapsto N_s$ is discontinuous at $s=r$,
\item[(3)] $b_1(B_r(\omega))=b_2(B_r(\omega))=0$.
\end{enumerate}
\bl{hiroshima1}
For every $(x,\omega,\omega_1,\omega_2)\in \OO^c$ we have
$$
\lim_{\varepsilon\to 0} \left| e^{\int_0^{T_t+} \log (-\chi_\varepsilon
(\uo(B_s,-\c_{s-})))dN_s}\right|=0.
$$
\el
\proof
We have $|e^{\int_0^{T_t+} \log (-\chi_\varepsilon (\uo(B_s,-\c_{s-})))dN_s} |\leq e^{ \int_0^{T_t+}
\log (W(B_s)+\varepsilon)dN_s}$. Observe that
$$
\int_0^{T_t+}\log (W(B_s)+\varepsilon)dN_s=\sum_{j=1}^n\log (W(B_{r_j})+\varepsilon), \quad r_1,...,r_n\in (0,T_t].
$$
Since $(x,\omega,\omega_1,\omega_2)\in \OO^c$, there exists an $r_i$ such that $b_1(B_{r_i}(\omega))
=b_2(B_{r_i}(\omega))=0$. Then
$$
\int_0^{T_t+}\log(W(B_s)+\varepsilon)dN_s= \sum_{j\not=i}^n\log(W(B_{r_j})+\varepsilon)+\log\varepsilon,
$$
and $e^{\int_0^{T_t+} \log (W(B_s)+\varepsilon)dN_s} \leq e^{\sum_{j\not=i}^n \log (W(B_{r_j})+\varepsilon)}
e^{\log\varepsilon}$. Thus
$
\lim_{\varepsilon\to0} |e^{\int_0^{T_t+} \log (W(B_s)+\varepsilon)dN_s} |=0,
$
and the lemma follows.
\qed

\bt{main2}
{\rm (\textbf{Feynman-Kac formula: magnetic field with zeros})}
Let Assumptions \ref{fumio4}  and   \ref{70} hold, and suppose that $V$ is relatively bounded with
respect to $\sqrt{-\Delta +m^2}$ with a relative bound strictly less than 1. Let $\OO$ be given by
\kak{oo}. Then
\eq{ko16new}
(f, e^{-t(H+V)}g) = \AAA \int_{\RR^3} dx \MMMM\lkkk e^{T_t}\ov{f(q_0)}
g(q_t)e^{\SSS}\one_{\OO} \rkkk.
\en
\et
\proof
Put $V=0$ and fix $\varepsilon>0$. We can show that the functional integral representation of
$\HHHH$ is given by \kak{MAIN} with $\SSS$ replaced by $\SSS_{\rm A} + \SSS_{\rm S} (\varepsilon)$
with
\begin{eqnarray}
\SSS_{\rm S} (\varepsilon) =
-\int_0^{T_t} \ud(B_s,\c_s) ds +\int_0^{T_t+} \log \lk -\chi_\varepsilon
(\uo(B_s, -\c_{s-})) \rk dN_s.
\end{eqnarray}
That is,
\eq{hh123}
(f, e^{-t\HHHH }g) = \AAA \int_{\RR^3} dx \MMMM\lkkk e^{T_t}\ov{f(q_0)}
g(q_t)e^{\SSS_{\rm A}+\SSS_{\rm S} (\varepsilon)} \rkkk.
\en
Take the limit $\varepsilon\downarrow0$ on both sides above. This gives
\eq{conv1}
\lim_{\varepsilon\downarrow 0} \exp\left(-t \HH_\varepsilon \right) = \exp\left(-t H \right)
\en
in strong sense, obtained in the same way as Lemma \ref{n2}. On the other hand, by the Lebesgue
dominated convergence theorem it follows that
$$
\lim_{\varepsilon\downarrow 0} \int_{\RR^3} dx \MMMM\lkkk e^{T_t}\ov{f(q_0)}
g(q_t)e^{\SSS_{\rm A}+\SSS_{\rm S} (\varepsilon)} \rkkk= \int_{\RR^3} dx \MMMM\lkkk
\lim_{\varepsilon\downarrow 0}e^{T_t}\ov{f(q_0)} g(q_t)e^{\SSS_{\rm A}+\SSS_{\rm S} (\varepsilon)}
\rkkk.
$$
By Lemma \ref{hiroshima1} we find that $\lim_{\varepsilon\to 0} \SSS_{\rm S} (\varepsilon) = 0$
on $\OO$ and hence
$$
\lim_{\varepsilon\downarrow 0} e^{\SSS_{\rm A}+\SSS_{\rm S} (\varepsilon)} =
\lim_{\varepsilon\downarrow 0} e^{\SSS_{\rm A}+\SSS_{\rm S} (\varepsilon)}\one_{\OO} +
\lim_{\varepsilon\downarrow 0} e^{\SSS_{\rm A}+\SSS_{\rm S} (\varepsilon)}\one_{\OO^c}
= e^{\SSS_{\rm A}+\SSS_{\rm S} }\one_\OO.
$$
Next suppose that $V\in L^\infty(\BR)\cap C(\BR)$. In this case we can show the theorem in the same
way as in Step 3 in the proof of Theorem \ref{main}. Furthermore, the theorem holds for the required
$V$ in the same way as in Steps 4 and 5 above.
\qed

A diamagnetic inequality follows immediately from Theorem \ref{main2}. Recall that $\HH_{b_0}$ is
defined by $H$ with $b$ replaced by $b_0=(\sqrt{b_1^2+b_2^2},0,b_3)$ and $a$ by zero, respectively.
\bc{g}
{\rm (\textbf{Energy comparison inequality})}
Under the assumptions of Theorem~\ref{main2} we have that
\eq{s}
|(f,e^{-t(\HH+V)}g)| \leq (|f|,e^{-t(\HH_{b_0}+V)}|g|).
\en
In particular, it follows that $\is{\HH_{b_0}+V}\leq \is{H+V}$.
\ec

\section{Fall-off of bound states}
\subsection{Martingale properties: non-relativistic case}
In this section we prove the decay properties of bound states of relativistic Schr\"odinger
operators with spin by means of the Feynman-Kac formula derived in the previous section. For simplicity
we assume throughout that
\eq{hh32100}
\EE^{x}_{P} \lkkk\int_0^t {|\log W(B_s)|}ds\rkkk <\infty,\quad {\rm a.e.} \; x\in\RR^3,
\en
and
\eq{hh321000}
\EE^{x,0}_{P\times \nu} \lkkk\int_0^{T_t} {|\log W(B_s)|}ds\rkkk <\infty,\quad {\rm a.e.} \; x\in\RR^3,
\en
i.e., the measure of $\OO^c$ in \kak{oo} is zero.

We  first consider the non-relativistic case. Let $\HHHHH$ be the Hamiltonian defined by
\eq{n111}
\HHHHH=h+V,
\en
where $h$ is given by \kak{ss1}.
Let $\SSSS$ be defined by the exponent $\SSS$ with the subordinator $T_t$ replaced by the non-random time $t$.
If Assumptions \ref{fumio4} and \ref{70} hold and  $V$ is relatively bounded with respect to $-\Delta$ with a 
relative bound strictly smaller than $1$, then $h+V$ is self-adjoint on $D(-\Delta)$. Then the Feynman-Kac 
formula of $(f, e^{-t(h+V)} g)$ is 
\eq{kon16}
(f, e^{-t(h + V)}g)=\AAA\int_{\RR^3} dx \EE^{x,\alpha}_{P\times\mu}\lkkk e^{t}
\ov{f(B_0,\c_0)} g(B_t,\c_t)e^{\SSSS} \rkkk,
\en
where  the exponent $\SSSS={\SSSS}_V +{\SSSS}_{\rm A} +{\SSSS}_{\rm S}$ is given by
\begin{eqnarray*}
&&
{\SSSS}_V  =-\int_0^t V(B_{s}) ds, \\
&&
{\SSSS}_{\rm A}  =-i\int_0^t a(B_s)\circ dB_s,\\
&&
{\SSSS}_{\rm S} = -\int_0^t
\ud
(B_s,\c_s) ds + \int_0^{t+}
\log\lk -\uo(B_s,- \c_{s-})\rk dN_s.
\end{eqnarray*}
Let $\gr$ be a bound state such that
$\HHHHH \gr=E\gr$ with $E\in\RR$.
We consider the spatial decay of $|\gr(x,(-1)^\alpha)|$, i.e., its behavior for large $|x|$.

Let
$\SSSS(x,\alpha)
={\SSSS}_V(x) +{\SSSS}_{\rm A}(x) +{\SSSS}_{\rm S}(x,\alpha)$
be  given by $\SSSS$ with $B_s$ and $N_s$ replaced by $B_s+x$ and $N_s+\alpha$, respectively:
\begin{eqnarray*}
&&
{\SSSS}_V (x)
 =-\int_0^t V(B_{s}+x) ds, \\
&&
{\SSSS}_{\rm A}  (x)
=-i\int_0^t a(B_s+x)\circ dB_s,\\
&&
{\SSSS}_{\rm S}(x,\alpha) = -\int_0^t
\ud
(B_s+x,(-1)^\alpha\c_s) ds + \int_0^{t+}
\log\lk -\uo(B_s+x,-(-1)^\alpha  \c_{s-})\rk dN_s.
\end{eqnarray*}
Define the stochastic process $(M_t(x,\alpha))_{t\geq0}$ by
$$M_t(x,\alpha)=e^{t(E+1)}e^{\SSSS(x,\alpha)}\gr(B_t+x,(-1)^\alpha \c_{t}),
\quad t\geq0,$$
and the filtration
$$\mmmm_t=\sigma((B_r,\c_{r}), 0\leq r\leq t),\quad t\geq0.$$
Note that
$e^{-t(\HHHHH-E)}\gr=\gr$ and
then
\eq{naka}
\EE^{x,\alpha}_{P\times\mu}[M_t(0,0)] =\EE^{0,0}_{P\times\mu}[M_t(x,\alpha)] =\gr(x,(-1)^\alpha)
\en
by \kak{kon16}.
\bl{martingale1}
The stochastic process $(M_t(x,\alpha))_{t\geq0}$ is a martingale with respect to $\pro \mmmm$, i.e., 
$\EE_{P\times\mu}^{0,0}[M_t(x,\alpha)|\mmmm_s]=M_s(x,\alpha)$ for $t\geq s$.
\el
\proof
We prove the case when $(x,\alpha)=(0,0)$ for notational simplicity, the proof for $(x,\alpha)\not=(0,0)$ 
is similar. Let $\SSSS([u,v])$ be defined by $\SSSS$ with the integration domain in $\int_0^t\cdots$ 
replaced by $\int_u^v\cdots$. Write $M_t=M_t(0,0)$. We see that
\begin{eqnarray*}
\EE^{0,0}_{P\times\mu} [M_t|\mmmm_s] = e^{t(E+1)} e^{\SSSS([0,s])} \EE^{0,0}_{P\times\mu}
\left[e^{\SSSS([s,t])} \gr(B_t,\c_t) |\mmmm_s\right].
\end{eqnarray*}
By the Markov property of the $\BR\times\zz$-valued stochastic process $(B_t,N_t)_{t\geq0}$ we have
\begin{eqnarray}
\lefteqn{
\EE^{0,0}_{P\times\mu} \lkkk e^{\SSSS([s,t])} \gr(B_t,\c_t) |\mmmm_s\rkkk \non }\\
&&=
\EE_{P\times\mu}^{B_s,N_s}\left[e^{-\int_0^{t-s} V(B_r)dr} e^{-i\int_0^{t-s} a(B_r)\circ dB_r}
e^{\int_0^{t-s} \ud(B_r,\c_r)dr} e^{K} \gr(B_{t-s},\c_{t-s}) \right].
\label{tachi}
\end{eqnarray}
The off-diagonal part $K$ in \kak{tachi} is
\begin{align*}
K
&=
\sum_{s<u\leq  t\atop N_{(u-s)+}\not= N_{(u-s)-}} \log(-\uo (B_{u-s}, -\c_{(u-s)-})) \\
&=
\sum_{0< r\leq  t-s\atop N_{r+}\not= N_{r-}} \log(-\uo (B_r, -\c_{r-})) =\int_0^{(t-s)+}
\log(-\uo (B_{r}, -\c_{r-})) dN_r.
\end{align*}
Hence we conclude that
\begin{eqnarray*}
\EE^{0,0}_{P\times\mu} [e^{\SSSS([s,t])}\gr(B_t,\c_t)|\mmmm_s] =
\EE^{B_s,N_s}_{P\times\mu} [e^{\SSSS([0,t-s])} \gr(B_{t-s},\c_{t-s})],
\end{eqnarray*}
which implies that
\begin{eqnarray*}
\EE^{0,0}_{P\times\mu}[M_t|\mmmm_s] = e^{s(E+1)} e^{\SSSS([0,s])} \EE^{B_s,N_s}_{P\times\mu}[M_{t-s}]
=M_s.
\end{eqnarray*}
Thus the lemma follows.
\qed

\subsection{Martingale properties: relativistic case}
Next we discuss the relativistic case $\HH+V$. Let $\gr$ be a bound state of $\HH+V$ such that
\eq{naga}
(\HH+V)\gr=E\gr
\en
for $E\in\RR$. We use the same notation $\gr$ as for the non-relativistic case. Consider the stochastic
process 
\eq{m1.1}
Y_t =e^{tE}e^{T_t} e^{\SSS}\gr(q_t),\quad t\geq0.
\en
Furthermore, we define
\eq{m1.2}
Y_t(x,\alpha) =e^{tE}e^{T_t} e^{\SSS(x,\alpha)}\gr(q_t(x,\alpha)),\quad t\geq0,
\en
where
$q_t(x,\alpha)=(B_{T_t}+x,(-1)^\alpha \theta_{T_t})$ and $\SSS(x,\alpha)=\SSS_V(x) +\SSS_{\rm A}(x) +
\SSS_{\rm S}(x,\alpha)$ is given by
\begin{align}
&
\label{london1-1}
\SSS_V  =-\int_0^t V(B_{T_s}+x) ds, \\
&
\label{london2-1}
\SSS_{\rm A}  =-i\int_0^{T_t } a(B_s+x)\circ dB_s,\\
&
\label{london3-1}
\SSS_{\rm S} = -\int_0^{T_t }
\ud
(B_s+x,(-1)^\alpha \c_s) ds + \int_0^{T_t+} \log\lk -\uo(B_s+x,- (-1)^\alpha\c_{s-})\rk dN_s.
\end{align}
Then
\eq{tamo}
\MMMM[Y_t]=\MMMMM[Y_t(x,\alpha)]=\gr(x,(-1)^\alpha).
\en
We introduce a filtration under which $\pro Y$ is a  martingale. Define $Y_t(\omega)$ and $Y_t(x,\alpha,\omega)$
for every $\omega\in \Omega_\nu$ by $Y_t$ and $Y_t(x,\alpha)$, respectively, and with subordinator $T_t$ replaced 
by the number $T_t(\omega)\geq 0$. Let
\eq{d3}
\fff_t^{(1)}(\omega)=\sigma((B_r, N_r), 0\leq r\leq T_t(\omega)) \in \fff_P\times\fff_\mu
\en
for $\omega\in \Omega_\nu$, and define
\eq{d1}
\fff_t^{(1)} =\lkk \left. \bigcup_{\omega\in\Omega_\nu}(A(\omega),\omega) \right|A(\omega)\in \fff_t^{(1)}(\omega)\rkk
\subset \fff_P\times\fff_\mu\times \fff_\nu.
\en
We also define
\eq{d2}
\fff_t^{(2)} = \lkk \left. \bigcup_{\omega\in\Omega_P\times\Omega_N} (\omega, B(\omega))
\right|B(\omega)\in \sigma(T_r, 0\leq r\leq t)\rkk \subset \fff_P\times\fff_\mu\times \fff_\nu.
\en
We see that $\fff_t^{(1)}$ and $\fff_t^{(2)}$ are sub-$\s$-fields of $\fff_P\times\fff_\mu\times \fff_\nu$.
Write
\eq{m4}
\fff_t=\fff_t^{(1)}\cap \fff_t^{(2)},\quad t\geq0.
\en
The conditional expectation $\MMMMM[Y_t(x,\alpha)|\fff_t^{(1)}]= 
\MMMMM[Y_t(x,\alpha)|\fff_t^{(1)}](\cdot,\cdot,\cdot)$ is a stochastic process on $\Omega_P\times\Omega_N\times\Omega_\nu$.
\bl{basic}
We have $\MMMMM[Y_t(x,\alpha)|\fff_t^{(1)}] (\cdot,\cdot,\omega) = \EE^{0,0}_{P\times\mu}[Y_t(x,\alpha,\omega)
|\fff_t^{(1)}(\omega)](\cdot,\cdot)$ for all $\omega\in\Omega_\nu$.
\el
\proof
Let $A=\bigcup_{\omega\in\Omega_\nu}(A(\omega),\omega)$ with $A(\omega)\in \fff_t^{(1)}(\omega)$.
Then
\begin{eqnarray*}
\MMMMM[1_AY_t(x,\alpha)]
&=&
\int _{\Omega_\nu}d\nu(\omega)
\EE^{0,0}_{P\times\mu}[1_{A(\omega)}Y_t(x,\alpha,\omega)]\\
&=&
\int _{\Omega_\nu}d\nu(\omega)
\EE^{0,0}_{P\times\mu}\left[1_{A(\omega)}(\cdot,\cdot)
\EE^{0,0}_{P\times\mu}\left[Y_t(x,\alpha, \omega)|\fff_t^{(1)}(\omega)\right](\cdot,\cdot)\right].
\end{eqnarray*}
On the other hand, we have
\eqn
\MMMMM[1_AY_t(x,\alpha)]
&=&
\MMMMM\left[1_A\MMMMM[Y_t(x,\alpha)|\fff_t^{(1)}]\right]\\
&=&
\int _{\Omega_\nu}d\nu(\omega) \EE^{0,0}_{P\times\mu}\left[1_{A(\omega)}
(\cdot,\cdot)\MMMMM\left[Y_t(x,\alpha)|\fff_t^{(1)}\right](\cdot,\cdot,\omega)\right].
\enn
A comparison of the two sides above completes the proof.
\qed
\bl{martingale2}
The stochastic process $(Y_t({x,\alpha}))_{t\geq0}$ is a martingale with respect to  $\pro \fff$, i.e.,
$\MMMMM[Y_t(x,\alpha)|\fff_s]=Y_s(x,\alpha)$ for $t\geq s$.
\el
\proof
We prove the case when $(x,\alpha)=(0,0)$ to keep the notation simple, the proof for $(x,\alpha)\not=(0,0)$ 
is again similar.

Note that $\MMMMM[Y_t|\fff_s]=\MMMMM[Y_t|\fff_s^{(1)}\cap \fff_s^{(2)}] 
=\MMMMM[\MMMMM[Y_t|\fff_s^{(1)}]|\fff_s^{(2)}]$. We first compute 
$\EE_{P\times\mu}^{0,0}[Y_t(\omega)|\fff_s^{(1)}(\omega)]$. Write
\begin{eqnarray*}
\SSS([u,v])
=
&&
-\int_u^v  V(B_{T_r})dr -i\int_{T_u} ^{T_v} a(B_r)\circ dB_r\\
&&
-\int_{T_u} ^{T_v}\ud (B_r,\c_r)dr +\int_{T_u}^{T_v+} \log(-\uo (B_r,-\c_{r-}))dN_r
\end{eqnarray*}
and, for every $\omega\in\Omega_\nu$
\begin{eqnarray*}
\SSS([u,v],\omega)
=
&&-\int_u^v  V(B_{T_r(\omega)})dr -i\int_{T_u(\omega)} ^{T_v(\omega)} a(B_r)\circ dB_r\\
&&
-\int_{T_u(\omega)} ^{T_v(\omega)}\ud (B_r,\c_r)dr +\int_{T_u(\omega)}^{T_v(\omega)+}
\log(-\uo (B_r,-\c_{r-}))dN_r
\end{eqnarray*}
and $q_t(\omega)=(B_{T_t(\omega)}, \c_{T_t(\omega)})$, $t\geq0$. Since $T_t(\omega)$ is non-random, 
we see in a similar way to the non-relativistic case that
\begin{eqnarray*}
\lefteqn{
\EE^{0,0}_{P\times\mu}[Y_t(\omega)|\fff_s^{(1)}(\omega)]} \\
&&
=
e^{tE}e^{T_t(\omega)} e^{\SSS([0,s],\omega)} \EE^{0,0}_{P\times\mu}[e^{\SSS([s,t],\omega)}\gr(q_t(\omega))
|\fff_s^{(1)}(\omega)]\\
&&
=
e^{tE}e^{T_t(\omega)} e^{\SSS([0,s],\omega)}\EE^{B_{T_s(\omega)},N_{T_s(\omega)}}_{P\times\mu}
\left[e^{-\int_s^t V(B_{T_r(\omega)-T_s(\omega)})dr} e^{-i\int_{T_s(\omega)} ^{T_t(\omega)}
a(B_{r-T_s(\omega)})\circ dB_r}\right.\\
&&\hspace{2cm}
\times\left.
e^{-\int_{T_s(\omega)}^ {T_t(\omega)} \ud (B_{r-T_s(\omega)},\c_{r-T_s(\omega)})dr}
e^{\int_{T_s(\omega)}^{T_t(\omega)+}\log(-\uo (B_{r-T_s(\omega)},-\c_{(r-T_s(\omega))-}))
dN_r}\right.\\
&&\hspace{7cm}
\left.\times \gr(B_{T_t(\omega)-T_s(\omega)}, \c_{T_t(\omega)-T_s(\omega)})\frac{}{}\right].
\end{eqnarray*}
Hence by Lemma \ref{basic} we have
\eqn
\MMMMM[Y_t|\fff_s^{(1)}] = e^{tE}e^{T_s}e^{\SSS([0,s])}Z_{t,s},
\enn
where
\eqn
&&
Z_{t,s}=e^{T_t-T_s} \EE^{B_{T_s},N_{T_s}}_{P\times\mu} \left[e^{-\int_s^t V(B_{T_r-T_s})dr}
e^{-i\int_{T_s} ^{T_t}a(B_{r-T_s})\circ dB_r}\right.\\
&&
\hspace{2cm}
\left.
\times e^{-\int_{T_s}^ {T_t} \ud (B_{r-T_s},\c_{r-T_s})dr} e^{\int_{T_s}^{T_t+}
\log(-\uo (B_{r-T_s},-\c_{(r-T_s)-})) dN_r} \gr(B_{T_t-T_s}, \c_{T_t-T_s})\right].
\enn
Here $Z_{t,s}$ is given by
\eqn
&&
e^{u-v} \EE^{B_{u},N_{v}}_{P\times\mu} \left[e^{-\int_s^t V(B_{T_r-u})dr}e^{-i\int_{v} ^{u}
a(B_{r-v})\circ dB_r}\right.\\
&&
\hspace{2cm}
\left.
\times e^{-\int_{v}^ {u} \ud (B_{r-v},\c_{r-v})dr} e^{\int_{v}^{u+} 
\log(-\uo (B_{r-v},-\c_{(r-v)-})) dN_r} \gr(B_{u-v}, \c_{u-v})\right]
\enn
evaluated at $u=T_t$ and $v=T_s$. Take the conditional expectation of the right hand side above 
with respect to $\fff_s^{(2)}$. We note that
\eq{ce1}
\MMMMM[f|\fff_s^{(2)}](\omega_1,\omega_2,\cdot) = \EE_\nu^0[f(\omega_1,\omega_2,\cdot)|\ms N_s](\cdot),
\en
where $\ms N_s=\s(T_r,0\leq r\leq s)$. Since $e^{tE}e^{T_s}e^{\SSS([0,s])}$ is measurable with respect 
to $\fff_s^{(2)}$, by \kak{ce1} we consider the conditional expectation of $Z_{t,s}$ giving
\begin{align*}
&\MMMMM \left[Z_{t,s}|\fff_s^{(2)}\right]\\
&=
\EE_\nu^0\lkkk e^{T_t-T_s} \EE^{B_{T_s},N_{T_s}}_{P\times\mu} \lkkk
e^{-\int_s^t V(B_{T_r-T_s})dr} e^{-i\int_0 ^{T_t-T_s} a(B_{r})\circ dB_r} \right. \right.
\\
&
\left.
\left.
\hspace{1cm}
\left.\times e^{-\int_0^ {T_t-T_s} \ud (B_{r},\c_{r})dr} e^{\int_0^{(T_t-T_s)+} \log(-\uo (B_{r},-\c_{r-}))
dN_r} \gr(B_{T_t-T_s}, \c_{T_t-T_s}) \rkkk \right|\ms N_s\rkkk,
\end{align*}
where we used the Markov property of $((B_t,N_t))_{t\geq0}$. By the Markov property of $\pro T$ we have
\begin{align*}
&=
\EE_\nu^{T_s}\lkkk e^{T_{t-s}-T_0} \EE^{B_{T_0},N_{T_0}}_{P\times\mu} \lkkk e^{-\int_s^t V(B_{T_{r-s}-T_0})dr}
e^{-i\int_0 ^{T_{t-s}-T_0} a(B_{r})\circ dB_r} \right. \right.
\\
&
\left.
\left.
\hspace{1cm}
\times e^{-\int_0^ {T_{t-s}-T_0} \ud (B_{r},\c_{r})dr} e^{\int_0^{(T_{t-s}-T_0)+} \log(-\uo (B_{r},-\c_{r-}))
dN_r} \gr(B_{T_{t-s}-T_0}, \c_{T_{t-s}-T_0}) \rkkk \rkkk.
\end{align*}
Since $\EE_\nu^u[f(T_\cdot)]= \EE_\nu^0[f(T_\cdot+u)]$, we see that
\begin{align*}
&=
\EE_\nu^0\lkkk e^{T_{t-s}-T_0} \EE^{B_{T_0+u},N_{T_0+u}}_{P\times\mu} \lkkk
e^{-\int_s^t V(B_{T_{r-s}-T_0})dr} e^{-i\int_0 ^{T_{t-s}-T_0} a(B_{r})\circ dB_r} \right. \right.
\\
&
\left.
\left.
\hspace{1cm}
\left.
\times e^{-\int_0^ {T_{t-s}-T_0} \ud (B_{r},\c_{r})dr} e^{\int_0^{(T_{t-s}-T_0)+}
\log(-\uo (B_{r},-\c_{r-})) dN_r} \gr(B_{T_{t-s}-T_0}, \c_{T_{t-s}-T_0}) \rkkk
\rkkk\right\lceil_{u=T_s}
\\
&=
\EEN^{B_{T_s},N_{T_s},0}
\lkkk
e^{T_{t-s}}
e^{-\int_0^{t-s} V(B_{T_r})dr}
e^{-i\int_0 ^{T_{t-s}}
a(B_{r})\circ dB_r}\right.\\
&\hspace{2cm}
\left.
\times
e^{-\int_0^ {T_{t-s}}
\ud (B_{r},\c_{r})dr}
e^{\int_0^{T_{t-s}+}
\log(-\uo (B_{r},-\c_{r-}))
dN_r}
\gr(q_{t-s})
\rkkk\\
&=
\lk
e^{-(t-s)\HHH}\gr\rk
(q_s).
\end{align*}
Hence  we conclude that
$$
\MMMMM[Y_t|\fff_s]=e^{sE}e^{T_s} e^{\SSS([0,s])} (e^{-(t-s)(\HHH-E)}\gr)(q_s)=Y_s
$$
and the lemma follows.
\qed

\subsection{Upper estimates on bound states}
We will use the following conditions.
\begin{assumption}
\label{strong}
The following properties hold:
\begin{enumerate}
\item[(1)]
$b_3\in L^\infty$ and $W=\sqrt{b_1^2+b_2^2}\in L^\infty$.
\item[(2)]
With $\mass=\|b_3\|_\infty+\|W\|_\infty$, we have $\mass<m^2/2$.
\item[(3)]
$V$ is of relativistic Kato-class, i.e.,
\eq{naga1}
\lim_{t\downarrow 0} \sup_{x\in\BR}\EE^{x,0}_{P\times\nu}\left[\int_0^t V(B_{T_r})dr\right]=0.
\en
\end{enumerate}
\end{assumption}
\bl{basic2}
If Assumption \ref{strong} holds, then $\gr\in L^\infty(\BR)$ and
\eq{m10}
|\gr(x,(-1)^\alpha)|
\leq
\EE_{P\times\nu}
^{0,0}
\left[
e^{(t\wedge \tau) E}
e^{-\int_0^{t\wedge \tau} V(B_{T_r}+x)dr}
e^{\frac{1}{2}T_{t\wedge \tau} \mass}
\right]
\|\gr\|
\en
for every stopping time $\tau$  with respect to $(\fff_s)_{s\geq0}$ and $t\geq0$.
\el
\proof
Notice that
$
\gr(x,(-1)^\alpha)=
\MMMM[Y_t]
$
for every $t$.
Then Schwarz inequality yields that
\eqn
&&
|\gr(x,(-1)^\alpha)|\\
&&\leq
e^{tE}
\EEN^{x,\alpha,0}
\left[
e^{2T_t}
e^{-2\int_0^t V(B_{T_r})dr}
e^{ \int_0^{T_t} |b_3(B_r)|dr}
e^{ \int_0^{T_t+}
\log W(B_r) dN_r}
\right]^\han
\MMMM\left[
|\gr(q_t)|
\right]^\han\\
&&
\leq
e^{tE}
\lk
\EE^{0,0}_{P\times\nu}
[
e^{-2\int_0^t V(B_{T_r}+x)dr}
e^{ {T_t}\mass}]\rk^\han
\lk
\MMMM
[|\gr(q_t)|^2]\rk^\han.
\enn
Here we used that
$\EE_\mu^0[e^{N_{T_t}\log W}]=e^{T_t(W-1)}$.
Note that
$$
\MMMM
[|\gr(q_t)|^2]=
\int_0^\infty  ds p_t(s)
\int_\BR
\Pi_s(y)dy
\sum_{n=0}^\infty |
\gr(x+y, (-1)^{\alpha+n})|^2  \frac{s^n}{n!}e^{-s},$$
where $\d p_t(s)=\frac{te^{tm}}{\sqrt{2\pi s^3}}
e^{-\half (\frac{t^2}{s}+m^2 s)}\one_{[0,\infty)}(s)$  denotes the distribution of subordinator $T_t$.
Since
$$\sum_{n=0}^\infty
|\gr(x+y, (-1)^{\alpha+n})|^2  \frac{s^n}{n!}e^{-s}
\leq
|\gr(x+y, 1)|^2+
|\gr(x+y, -1)|^2,$$
we obtain
\eqn
\lefteqn{
\MMMM
[|\gr(q_t)|^2]}\\
&&\leq
\int_\BR dy
\int_0^\infty ds p_t(s) \Pi_s(y)
(|\gr(x+y, 1)|^2+
|\gr(x+y, -1)|^2)\\
&&=
\int_\BR dy
(|\gr(x+y, 1)|^2+
|\gr(x+y, -1)|^2)\pi^{-2}\frac{te^{mt}}
{(|y|^2+t^2)^2}
\int _0^\infty
\xi
e^{-(\xi+\frac{m^2(|y|^2+t^2)/4}{\xi})}d\xi\\
&&\leq  C_t\|\gr\|_{L^2(\BR\times\zz)}^2
\enn
with a constant $C_t$. Furthermore,  let $m^2/(2\mass)>q>1$ and $\frac{1}{p}+\frac{1}{q}=1$.
Then we get
$$
\EE_{P\times\nu}^{x,0}
[
e^{-2\int_0^t V(B_{T_r})dr}
e^{{T_t}\mass}]
\leq
\lk
\EE_{P\times\nu}
^{x,0}
[
e^{-2p\int_0^t V(B_{T_r})dr}]\rk
^{1/p}
\lk
\EE_{P\times\nu}^{x,0}
[
e^{q{T_t}\mass}]\rk
^{1/q}.
$$
The first term at the right hand side above satisfies that
\eq{kobu}
\sup_{x\in\BR}\lk
\EE^{x,0}_{P\times\nu}\left[
e^{-2p\int_0^t V(B_{T_r})dr}\right]\rk^{1/p}<\infty
\en
since $V$ is of relativistic Kato class, and
\eq{n20}
\EE^{x,0}_{P\times \nu}
[e^{q{T_t}\mass}]=
\EE^{0}_\nu
[e^{q{T_t}\mass}]=
\int_0^\infty {e^{qs\mass}} \frac{te^{mt}}{\sqrt{2\pi s^3}}
e^{-\frac{1}{2}(\frac{t^2}{s}+m^2s)}ds = {e^{+t(m-\sqrt{m^2-2q\mass})}}<\infty.
\en
Hence $\gr\in L^\infty(\BR)$.
Notice that
by the martingale property of $Y_t(x,\alpha)$,
\eq{m9}
\gr(x,(-1)^\alpha)=
\MMMMM[Y_{t\wedge\tau}(x,\alpha)]
\en
for every stopping time $\tau$ and $t\geq0$.
\kak{m10} follows from \kak{m9} and
\begin{eqnarray*}
|\gr(x,(-1)^\alpha)|
&\leq&
\EE^{0,0}_{P\times\nu}
\left[
e^{(t\wedge \tau)E}
e^{-\int_0^{t\wedge \tau} V(B_{T_r}+x)dr}
e^{T_{t\wedge \tau}\mass/2}
\right]
\|\gr\|.
\end{eqnarray*}
\qed

\subsection{Decay of bound states: the case $V\to\infty$}
In this subsection we show the spatial exponential decay
of bound states of $\HH+V$ at infinity.
\bl{stopping time}
Let $\tr=\inf\{t||B_{T_t}|>R\}$. Then
$\tr$ is a stopping time
with respect to the filtration $\pro \fff$.
\el
\proof
It suffices to show that $\{\tr\leq t\}\in \fff_t$.
Notice  that
$$\{\tr\leq t\}=
\bigcup_{\omega\in\Omega_\nu}(A(\omega),\omega),$$
where $A(\omega)=\{\omega'\in \Omega_P|
\sup_{0\leq s\leq t}|B_{T_s(\omega)}(\omega')|> R\}
\in \fff_t^{(1)}(\omega)$.
Thus $\{\tr\leq t\}\in \fff_t^{(1)}$.
Moreover
$$\{\tr\leq t\}=
\bigcup_{\omega\in\Omega_P}(\omega, B(\omega)),$$
where $B(\omega)=\{\omega'\in \Omega_\nu|
\sup_{0\leq s\leq t}|B_{T_s(\omega')}(\omega)|>R\}$.
Therefore
$\{\tr\leq t\}\in\fff_t^{(2)}$ and hence
$\{\tr\leq t\}\in\fff_t$.
\qed

\bt{main3}
If Assumption \ref{strong} holds
and
\eq{n6}\lim_{|x|\to\infty}V(x)=\infty,
\en
then for every $a>0$ there exists $b>0$ such that
\eq{n14}
|\gr(x,(-1)^\alpha)|\leq b  e^{-a|x|}.
\en
\et
\proof
We have by Lemma \ref{basic2} that
$$|\gr(x,(-1)^\alpha)|\leq
\lk
\EE^{0,0}_{P\times\nu}
\lkkk
e^{2(t\wedge \tr) E}
e^{-2\int_0^{t\wedge \tr}
V(B_{T_r}+x) dr}
\rkkk\rk ^\han
\lk
\EE^{0,0}_{P\times\nu}
\lkkk
e^{\mass T_{t\wedge \tr}}
\rkkk \rk
^\han
\|\gr\|.$$
Let $W(x)=W_R(x)=\inf \{V(y)||x-y|<R\}$, and notice that
\eq{n21}
\lim_{|x|\to\infty}W(x)-E=\infty.
\en
In particular, we may assume that $W(x)-E>0$.
This gives
\eqn
\lefteqn{
\lk
\EE^{0,0}_{P\times\nu}
\lkkk
e^{2(t\wedge \tr) E}
e^{-2\int_0^{t\wedge \tr}
V(B_{T_r}+x) dr}
\rkkk\rk
^\han}\\
&&\leq
\lk
\EE^{0,0}_{P\times\nu}
\lkkk
e^{-2(t\wedge \tr) (W(x)-E)}
\rkkk \rk^\han\\
&&\leq
\lk
\EE^{0,0}_{P\times\nu}
\lkkk
1_{\{\tr<t\}}
e^{-2(t\wedge \tr) (W(x)-E)}
\rkkk\rk
^\han
+
\lk
\EE^{0,0}_{P\times\nu}
\lkkk
1_{\{\tr\geq t\}}
e^{-2(t\wedge \tr) (W(x)-E)}
\rkkk\rk
^\han\\
&&\leq
\lk
\EE^{0,0}_{P\times\nu}
\lkkk
1_{\{\tr<t\}}
\rkkk\rk
^\han
+e^{-t (W(x)-E)}.
\enn
We see that
\eq{cms}
\EE^{0,0}_{P\times\nu}
\lkkk
1_{\{\tr<t\}}
\rkkk
=
\EE^{0,0}_{P\times\nu}
\lkkk
1_{\{\sup_{0\leq s\leq t}|B_{T_s}|-R\geq0\}}
\rkkk
\leq
\EE^{0,0}_{P\times\nu}
\lkkk
e^{\alpha(\sup_{0\leq s\leq t}|B_{T_s}|-R)}
\rkkk
\en
for any $\alpha\geq0$.
 It can be shown that
$\lk
\EE^{0,0}_{P\times\nu}\lkkk
e^{\alpha\sup_{0\leq s\leq t}|B_{T_s}|}
\rkkk\rk
^\han
\leq C_1 e^{C_2 t}$ for sufficiently small $\alpha$,
see \cite[Proposition II.5]{CMS90}. Hence
$\lk
\EE^{0,0}_{P\times\nu}[1_{\{\tr\leq t\}}]\rk^\han
\leq e^{-\alpha R/2}C_1 e^{C_2 t}$, and
\eq{n11}
\lk
\EE^{0,0}_{P\times\nu}
\lkkk
e^{2(t\wedge \tr) E}
e^{-2\int_0^{t\wedge \tr}
V(B_{T_r}+x) dr}
\rkkk\rk^\han
\leq
e^{-t (W(x)-E)}
+
e^{-\alpha R/2}C_1 e^{C_2 t}.
\en
We also see that
\eqn
\EE^{0,0}_{P\times\nu}
\lkkk
e^{\mass T_{t\wedge \tr}}
\rkkk
&\leq&
\EE^{0,0}_{P\times\nu}
\lkkk
1_{\{t<\tr\}}
e^{\mass T_{t\wedge \tr}}
\rkkk
+\EE^{0,0}_{P\times\nu}
\lkkk
1_{\{t\geq \tr\}}
e^{\mass T_{t\wedge \tr}}
\rkkk\\
&\leq &
\EE^{0,0}_{P\times\nu}
\lkkk
e^{\mass T_t}
\rkkk
+\EE^{0,0}_{P\times\nu}
\lkkk
1_{\{t\geq \tr\}}
e^{\mass T_{\tr}}
\rkkk\\
&\leq &
2 \EE^{0}_{\nu}
\lkkk
e^{\mass T_t}
\rkkk,
\enn
where we used that $T_{\tr}\leq T_t$
for $\tr \leq t$.
Thus we have
\eq{n22}
\EE^{0,0}_{P\times\nu}
\lkkk
e^{\mass T_{t\wedge \tr}}
\rkkk^\han
\leq
\sqrt 2
e^{t(m-\sqrt{m^2-2\mass})/2}
\en
by \kak{n20}.
Hence  by \kak{n11} and \kak{n22},
\eq{n23}
|\gr(x,(-1)^\alpha)|\leq
\sqrt 2
\lk
e^{-t (W(x)-E)}
+
e^{-\alpha R/2}C_1 e^{C_2 t}
\rk
e^{t(m-\sqrt{m^2-2\mass})/2}
\|\gr\|.
\en
Notice  that
by inserting $R=p|x|$ with any $0<p<1$,
$W(x)-E=W_R(x)-E=W_{p|x|}(x)-E\to\infty$ as $|x|\to\infty$.
Thus
substituting $t=\delta|x|$ for sufficiently small $\delta >0$ and $R=p|x|$ with some $0<p<1$ in \kak{n23},
the theorem follows.
\qed

\subsection{Decay of bound states: the case $V\to0 $}
In this subsection we consider the case of potentials decaying to zero as $|x|\to\infty$.
\bt{main 4}
Let  Assumption \ref{strong} hold
and
suppose that
\eq{m6}\lim_{|x|\to\infty}V(x)=0.
\en
Also, assume
that
\eq{n34}
m-\sqrt{m^2-2\mass}<-2E.
\en
Then
there exist $a, b>0$ such that
\eq{n14-1}
|\gr(x,(-1)^\alpha)|\leq b  e^{-a|x|}.
\en
\et
\proof
Define
$
\tr=\tr(x)=\inf\{t\geq0 |\; |B_{T_t}+x|\leq R\}$.
Then $\tr$ is a stopping time,
which can be seen in the same way as
in Lemma \ref{stopping time}.
Thus
\begin{align*}
|\gr(x,(-1)^\alpha)|&
\leq
\lk
\EE^{0,0}_{P\times\nu}
\lkkk
e^{2(t\wedge \tr) E}
e^{-2\int_0^{t\wedge \tr}
V(B_{T_r}+x) dr}
\rkkk \rk
^\han
\lk
\EE^{0,0}_{P\times\nu}
\lkkk
e^{\mass T_{t\wedge \tr}}
\rkkk\rk ^\han\|\gr\|\\
&=
\lk
\EE^{x,0}_{P\times\nu}
\lkkk
e^{2(t\wedge \tr(0)) E}
e^{-2\int_0^{t\wedge \tr(0)}
V(B_{T_r}) dr}
\rkkk \rk
^\han
\lk
\EE^{x,0}_{P\times\nu}
\lkkk
e^{\mass T_{t\wedge \tr(0)}}
\rkkk\rk ^\han\|\gr\|.
\end{align*}
We rewrite $\tr(0)$ by $\tr$.
Let $\varepsilon>0$ be arbitrary. Then for sufficiently large $R$ it follows that
$\sup_{|x|>R}|V(x)|<\varepsilon$ by \kak{m6}, and
we see that
$|\int_0^{t\wedge \tr}V(B_{T_r}) dr|\leq (t\wedge \tr )\varepsilon$.
This gives
$$|\gr(x,(-1)^\alpha)|\leq
\lk
\EE^{x,0}_{P\times\nu}[e^{2(t\wedge\tr)(E+\varepsilon)}]\rk
^\han
\lk
\EE^{x,0}_{P\times\nu}
\lkkk
e^{\mass T_{t\wedge \tr}}
\rkkk\rk ^\han\|\gr\|.$$
Thus
\eqn
\EE^{x,0}_{P\times\nu}[e^{2(t\wedge\tr)(E+\varepsilon)}]
&=&
\EE^{x,0}_{P\times\nu}[1_{\{t\leq \tr\}}
e^{2t (E+\varepsilon)}]
+
\EE^{x,0}_{P\times\nu}[1_{\{t>\tr\}}
e^{2\tr (E+\varepsilon)}]\\
&\leq &
e^{2t (E+\varepsilon)}
+C_1e^{-m_\varepsilon|x|}
\enn
by making use of \cite[(II.29)(II.22) and (IV.3)]{CMS90} as above, where
$$m_\varepsilon=\lkk
\begin{array}{ll}
m & \mbox{if \, $2|E|>m$} \\
2\sqrt{m|E|-|E|^2} & \mbox{if \, $2|E|\leq m$}.
\end{array}
\right.$$
Also, notice
that
$$\EE^{x,0}_{P\times\nu}[e^{\mass T_{t\wedge \tr}}]
\leq
2e^{t(m-\sqrt{m^2-2\mass})}.$$
Therefore
\eq{n33}
|\gr(x,(-1)^\alpha)|
\leq
(e^{t (E+\epsilon)}
+C_1e^{-m_\epsilon|x|/2}
)
\sqrt2
e^{t(m-\sqrt{m^2-2\mass})/2}.
\en
On inserting  $t=\delta |x|$ with sufficiently small $\delta$,
the theorem follows from \kak{n34}.
\qed

\bigskip
\noindent {\bf Acknowledgments:}
FH acknowledges support of Grant-in-Aid for Science Research (B) 20340032 from JSPS and
Grant-in-Aid for Challenging Exploratory Research 22654018 from JSPS, and thanks the
hospitality of ICMS Edinburgh, where part of this work was done. JL also thanks ICMS
Edinburgh for a Research-in-Group grant allowing to complete the project of this paper.
We thank Toshimitsu Takaesu for pointing out typing errors in the first version of the
manuscript.

\end{document}